\RequirePackage{fix-cm}
\documentclass[twocolumn]{svjour3}

\smartqed  

\usepackage{graphicx}
\usepackage{gnuplot-lua-tikz}
\usepackage{hyperref}
\usepackage{lineno}
\usepackage{mathtext}
\usepackage{mathtools}
\usepackage{xcolor}

\usepackage[misc,geometry]{ifsym}

\definecolor{orcidlogocol}{HTML}{A6CE39}

\usepackage{etoolbox}
\newcommand*{\affaddr}[1]{#1} 
\newcommand*{\affmark}[1][*]{\textsuperscript{#1}}

\begin{document}

\title{The numerical scheme splitting along the coordinates in models of hydrocarbon migration
    based on Darcy flow concept and the mass conservation law\\
    \textcolor{red}{Preprint}}

\titlerunning{NSS-method in models of hydrocarbon migration based on Darcy flow concept}

\author{
    A.~Zhuravljov\affmark[1]$^{\;\textrm{\Letter}}$
    \and
    Z.~Lanetc\affmark[2,1]      
    \and    
    N.~Khoperskaya\affmark[1]
    \and
    S.~Rahman\affmark[2]
}

\authorrunning{A.~Zhuravljov  \and Z.~Lanetc \and N.~Khoperskaya \and S.~Rahman}

\institute{A.~Zhuravljov              
             \\a.s.zhuravljov@gmail.com
             \vspace{0.1cm}                   
               \\\at
               \affaddr{\affmark[1]Tyumen State University, Tyumen, 625003, Russia}\\
               \affaddr{\affmark[2]The University of New South Wales, Sydney, Australia}\\
}


\maketitle

\begin{abstract}
One of the most sophisticated and significant stages of basin modelling is hydrocarbon (HC) migration. In order to scrutinize this issue, it is expedient to utilize already available numerical reservoir simulation tools. Such tools are typically based on the Darcy flow model and are used to identify the structural features of the formation and estimate the production profile. Likewise, the migration of HCs can also be modelled allowing to clarify the possible location of HC deposits with higher accuracy. Despite all the advantages of this approach, there are some significant drawbacks, including the long computational time required to simulate the hydrodynamic process of HC migration. This issue cannot always be resolved by increasing computational power due to its technological scarcity. Thus, the authors suggest using a method of the numerical scheme splitting along the coordinates (further called NSS-method), allowing to significantly reduce the computational time when modelling HC migration. Explicit and implicit numerical models are created and scrutinized for the purpose of this research. The validity of these models is verified by their comparison with the available analytical solutions and by analyzing the stability of numerical schemes. As a result, the influence of NSS-method on the calculation accuracy was insignificant, allowing to decrease the computational time and number of time steps approximately by three orders of magnitude and $300$ times, respectively.

\end{abstract}

\section{Introduction}
\label{intro}

Geophysics and reservoir characterization are widely used in the extraction of HCs, allowing to identify the possible location of oil reservoirs and to clarify the structural features of a trap. However, a more in-depth analysis requires scrutinizing the geological processes in sedimentary basins. This can be achieved using basin modelling, consisting of an integrated framework of different phenomena, such as rock deposition and compaction, multiphase flow during HC migration and accumulation, as well as heat flow and phase composition analyses (Al-Hajeri et al 2009~\cite{al2009basin}). One of the crucial and most sophisticated processes in basin modelling is petro\-leum migration, which is responsible for the HC accumulation inside the reservoir (Allan 1989~\cite{allan1989model}; Catalan et al 1992~\cite{catalan1992experimental}; Thomas and Clouse 1995~\cite{thomas1995scaled}; Hantschel et al 2000~\cite{hantschel2000finite}; Schowalter 1979~\cite{schowalter1979mechanics}; Karlsen and Skeie 2006~~\cite{karlsen2006petroleum}).

According to  Hantschel and Kauerauf (2009~\cite{hantschel2009fundamentals}), the concepts of primary, secondary and tertiary migration are not subdivided in the basin modelling. Thus, the process of migration can be generalized and defined as the flow of HCs in the free pore space. Such phenomenon is caused by buoyancy force and capillary attraction. The buoyancy force leads to the gravity segregation allowing upward movement of HCs along a `carrier bed' from the source area to the trap. The existence of capillary imbibition leads to transport resistance and also results in the formation of the transition zone (Aziz and Settari 1979~\cite{aziz1979petroleum}; Jackson et al 2005~\cite{jackson2005prediction}). The migration intensity depends upon the formation inclination,  which is insignificant and typically varies in the range of several degrees (Chapman 2000~\cite{chapman2000petroleum}; Siddiqui and Lake 1992~\cite{siddiqui1992dynamic}; Bedrikovetsky et al 2001~\cite{bedrikovetsky2001secondary}). Therefore, the typical migration time is comparable with geological timescales (England et al 1987~\cite{england1987movement}).

Several factors complicate numerical simulation of HC migration, including its uncertain nature and extensive computing effort required. Different approaches exist to troubleshoot these issues, among them are map-based flowpath technique,  invasion percolation and hybrid methods (Hantschel and Kauerauf 2009~\cite{hantschel2009fundamentals}). These approaches have been developed as numerical algorithms using Darcy flow models are often too complex to be computed under acceptable times. Even considering small geological deposits, computational time appears too high and, thus, unreasonable. Notably, map-based flowpath technique and invasion percolation method are based on the crude flow approximations leading to a significantly higher error compared to the Darcy flow models (Luo 2011~\cite{luo2011simulation}; Carruthers 2003~\cite{carruthers2003modeling}).

Clearly, the conventional laws of multiphase fluid dynamics might be applied in order to investigate the phenomenon of HC migration (Aziz and Settari 1979~\cite{aziz1979petroleum}; Chen et al 2006~\cite{chen2006computational}; Lake 2007~\cite{Lake}). This can be accomplished using the well-developed instruments of numerical reservoir simulation. According to the above discussion, it is essential to investigate and develop methods allowing to reduce the required computational time when simulating the process of HC migration with the usage of the most accurate Darcy flow models. Therefore, the authors propose implementing NSS-method to accomplish this goal with an insignificant impact on the accuracy of the final results.

A brief description of the following sections is listed below. Section \ref{desc} presents the physico-mathematical model which consists of two conventional integral equations, corresponding initial and boundary conditions, and their numerical representations. Besides, geometric characteristics, fluid properties and a description of NSS-method are provided. Section~\ref{res} lists the obtained results, Sect.~\ref{dis} discusses the main findings of this study, while Sect.~\ref{con} summarizes the main conclusions based on these \mbox{findings}.

\section{Description of the calculation model\\and properties of fluids}
\label{desc}
\subsection{Physico-mathematical model}
\label{flu}

The primary research objective of the paper aims to optimize the conventional algorithms for numerical simulation of hydrodynamic aspects of oil migration (Hantschel and Kauerauf 2009~\cite{hantschel2009fundamentals}). The utilized physical model is represented by two-phase (water and oil) transport in porous media. Main assumptions neglect water, oil and reservoir compressibility, oversee tectonic deformations, and include the homogeneous reservoir rock and constant fluids viscosity. For the above phenomena, the physico-mathematical model corresponds to the following system of Equations (Aziz and Settari 1979~\cite{aziz1979petroleum}; Bear 2013~\cite{bear2013dynamics}; Dullien 2012~\cite{dullien2012porous}; Jamal et al 2006~\cite{jamal2006petroleum}; Lake 2007~\cite{Lake})

\begin{eqnarray}
\begin{gathered}
\label{eq:conserv_mass_1}
\int \limits_{V} \frac{\partial \phi S_{j}}{\partial t} d V + \oint \limits_{\Omega} \phi S_{j} \vec{\upsilon}_{j} d\vec{\Omega} = 0,
\end{gathered}
\end{eqnarray}

\begin{eqnarray}
\begin{gathered}
\label{eq:darcy}
\phi S_{j} \vec{\upsilon}_{j} = - \frac{k_{rj}\vec{\vec{k}}}{\mu_{j}} \vec{\nabla}\left(P_{j} - \rho_{j} \vec{g}\vec{x}\right),
\end{gathered}
\end{eqnarray}

\begin{eqnarray}
\begin{gathered}
\label{eq:sum_sat_2ph}
S_{w}+S_{o} = 1,
\end{gathered}
\end{eqnarray}

\begin{eqnarray}
\begin{gathered}
\label{eq:pressure_c}
P_{c} = P_{o} - P_{w},
\end{gathered}
\end{eqnarray}

\begin{eqnarray}
\begin{gathered}
\label{eq:initial_cond}
P_{w}\left(\vec{x}, 0\right) = P_{w}^{ini}, \; S_{w}\left(\vec{x}, 0\right) = S_{w}^{ini},
\end{gathered}
\end{eqnarray}

\begin{eqnarray}
\begin{gathered}
\label{eq:boundary_cond}
\phi S_{j} \vec{\upsilon}_{j} = 0 \; \text{on} \; \Gamma,
\end{gathered}
\end{eqnarray}
where $j=w\left(\text{water}\right),o\left(\text{oil}\right)$, $\phi$~--~porosity, $\rho$~--~density, $S$~--~saturation, $\vec{\upsilon}$~--~velocity, $k_{r}$~--~relative permeability, $\vec{\vec{k}}$~--~absolute permeability, $\mu$~--~dynamic viscosity, $P$~--~reservoir pressure, $t$~--~time, $V$~--~volume, $\Omega$~--~surface, $P_{c}$~--~the pressure difference is given by the capillary pressure.

Considering the above formulas, (\ref{eq:conserv_mass_1}) describes the conservation of mass law, while (\ref{eq:darcy})  represents empirical Darcy's equation. Additional algebraic relations (\ref{eq:initial_cond}) and (\ref{eq:boundary_cond}) are the initial and boundary conditions, respectively. Despite the impervious boundary is used, an oil influx is modelled by fixing the saturation value in the defined region. Such peculiar approach allows avoiding the contradiction between the amount of oil leaking and intensity of capillary to gravity driven imbibition.

For numerical simulation, it is convenient to convert the presented mathematical model~(\ref{eq:conserv_mass_1}~--~\ref{eq:pressure_c}) to the single Eq.~(\ref{eq:conserv_mass_2}), containing only one variable~$S_{w}$. An analogous mathematical formalism was firstly introduced by  Rapoport et al (1953~~\cite{rapoport1953properties}). Several following mathematical transformations are required to implement such an approach
\begin{eqnarray}
\begin{gathered}
\label{eq:darcy_difference}
\phi S_{w} \vec{\upsilon}_{w} \frac{\mu_{w}}{k_{w}} - \phi S_{o} \vec{\upsilon}_{o} \frac{\mu_{o}}{k_{o}} = \vec{\vec{k}} \vec{\nabla} \left(P_{c} +\left(\rho_{w}-\rho_{o}\right) \vec{g}\vec{x}\right),
\end{gathered}
\end{eqnarray}

\begin{eqnarray}
\begin{gathered}
\label{eq:darcy_sum_1}
\phi S_{w} \vec{\upsilon}_{w} + \phi S_{o} \vec{\upsilon}_{o}  = 0.
\end{gathered}
\end{eqnarray}
Equation (\ref{eq:darcy_difference}) is obtained by the combination of relations~(\ref{eq:darcy}), whereas Eq.~(\ref{eq:darcy_sum_1})~represents the effect of only capillary and gravity forces. According to Eq.~(\ref{eq:darcy_sum_1}), the only interpenetrating flow of phases is possible.

Expression~(\ref{eq:darcy_difference&sum_2}) is derived from the above relations and denotes the velocity of water. Substituting Eq.~(\ref{eq:darcy_difference&sum_2}) into Eq.~(\ref{eq:conserv_mass_1}) yields the target expression~(\ref{eq:conserv_mass_2}).
\begin{eqnarray}
\begin{gathered}
\label{eq:darcy_difference&sum_2}
\phi S_{w} \upsilon_{w} = \psi \vec{\vec{k}} \vec{\nabla} \left(P_{c} +\left(\rho_{w}-\rho_{o}\right) \vec{g}\vec{x}\right),
\end{gathered}
\end{eqnarray}
where function~$\psi$ specifies the relative mobility of phases to each other and is shown below:
\begin{eqnarray}
\begin{gathered}
\label{eq:diff_psi}
\psi = \frac{\frac{k_{rw}}{\mu_{w}}\frac{k_{ro}}{\mu_{o}}}{\frac{k_{rw}}{\mu_{w}}+\frac{k_{ro}}{\mu_{o}}}.
\end{gathered}
\end{eqnarray}

\begin{eqnarray}
\begin{split}
\label{eq:conserv_mass_2}
\int \limits_{V} \phi \frac{\partial S}{\partial t} d V + \oint \limits_{\Omega} \vec{\vec{k}} \psi P'_{c} \vec{\nabla} S & d\vec{\Omega}\;+ \\
+ \oint \limits_{\Omega} \left(\rho_{w}-\rho_{o}\right) & \vec{\vec{k}} \psi \vec{\nabla} \left(\vec{g}\vec{x} \right) d\vec{\Omega} = 0,
\end{split}
\end{eqnarray}

This research also aims to compare implicit and explicit numerical solution schemes. For this purpose, both schemes were implemented. Thus, Eqs.~(\ref{eq:conserv_mass_num_implicit}) and (\ref{eq:conserv_mass_num_explicit}) express the implicit and explicit finite difference representations  of Eq.~(\ref{eq:conserv_mass_2}), respectively (Aziz and Settari 1979~\cite{aziz1979petroleum}; Chen et al 2006~\cite{chen2006computational})
\begin{eqnarray}
\begin{gathered}
\label{eq:conserv_mass_num_implicit}
\frac{\alpha}{\Delta t^{n+1}}\Delta_{S}^{t} + \sum_{\Delta\Omega} \beta\nabla_{S}^{n+1} + \sum_{\Delta\Omega} \gamma\overline{\Delta}_{S}^{t} + \sum_{\Omega} \delta = 0,
\end{gathered}
\end{eqnarray}

\begin{eqnarray}
\begin{gathered}
\label{eq:conserv_mass_num_explicit}
\frac{\alpha}{\Delta t^{n+1}}\Delta_{S}^{t} + \sum_{\Delta\Omega} \beta\nabla_{S}^{n} + \sum_{\Delta\Omega} \delta = 0,
\end{gathered}
\end{eqnarray}
the boundary condition~(\ref{eq:boundary_cond}) is then transformed to the following:
\begin{eqnarray}
\begin{gathered}
\label{eq:boundary_cond_numerical}
\nabla_{S}^{m} = \overline{\Delta}_{S}^{t} = \delta = 0 \; \text{on} \; \Gamma.
\end{gathered}
\end{eqnarray}

To show the summation over all surface elements $\Delta\Omega$ of a grid block, operator~$\sum_{\Delta\Omega}$ is used. If the summation sign is not presented, the current grid block is characterized by the existing parameters. All equation coefficients  containing variables are presented in Eqs.~(\ref{eq:operators}), others might be found in Eqs.~(\ref{eq:coefficients}) (Ames 2014~\cite{ames2014numerical}; Smith 1985~\cite{smith1985numerical}).
\begin{eqnarray}
\begin{gathered}
\label{eq:operators}
\Delta_{S}^{t} = S^{n+1}-S^{n}, \; \overline{\Delta}_{S}^{t}= \overline{S}^{n+1}-\overline{S}^{n}, \\
\nabla_{S}^{m} = \frac{1}{L}\left(S_{+}^{m}-S_{-}^{m}\right),
\end{gathered}
\end{eqnarray}

\begin{eqnarray}
\begin{gathered}
\label{eq:coefficients}
\alpha = \phi V, \; \beta =  \overline{\psi} \varepsilon, \; \gamma = k \overline{\psi'} \epsilon, \; \delta = k \overline{\psi}\epsilon, 
\\
\varepsilon =  \overline{P'_{c}}\Delta\Omega, \; \epsilon = \left(\rho_{w}-\rho_{o}\right)g\frac{y_{-}-y_{+}}{L}\Delta\Omega,
\end{gathered}
\end{eqnarray}
where subscript signs '+' and '-' indicate the position of the finite difference block relative to the current surface element $\Delta\Omega$. 

The averaging operator in the cases of $\overline{\psi}$ and $\overline{\psi'}$ denotes the upstream weighting (Lake 2007~\cite{Lake}), in other cases the arithmetic mean is calculated. In both circumstances, averaging is conducted within grid blocks having a common current surface element  $\Delta\Omega$. The utilized approach of obtaining finite difference equations is widespread and described, for instance, by Aziz and Settari (1979~\cite{aziz1979petroleum}).

\subsection{Fluid properties and geometrical characteristics of the numerical models}
\label{1D}

\begin{figure*}[ht]
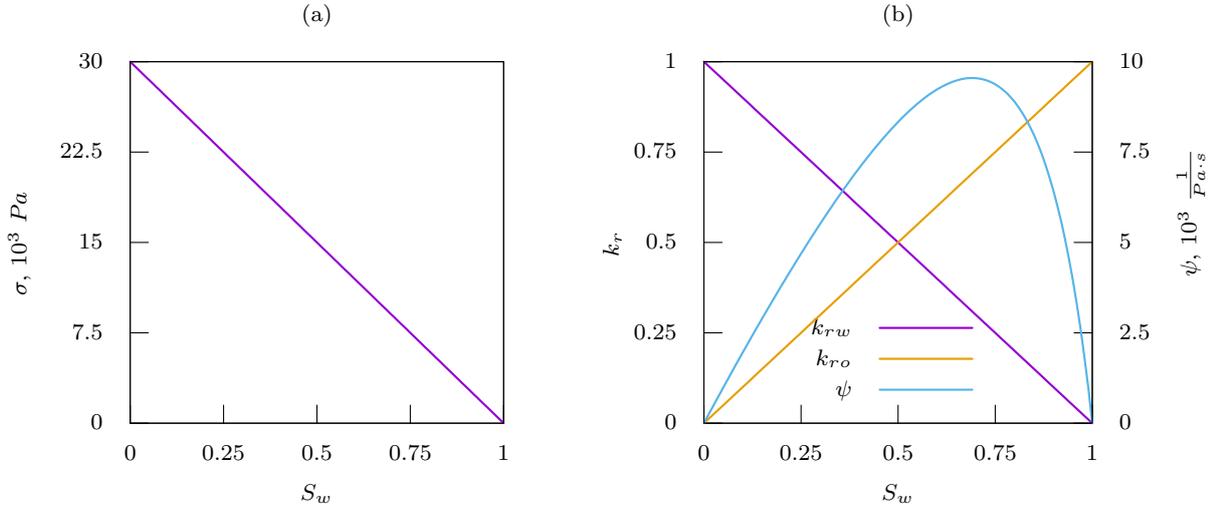
	
\centering
\input fig1.txt
\caption{Fluid properties: (\textbf{a}) capillary pressure function $\sigma$; (\textbf{b}) relative phase permeability curves $k_{r}$ and relative mobility $\psi$.}
\label{fig:1}
\end{figure*}

All constant fluid and rock parameters are listed in Table~\ref{tab:1}. The capillary pressure and two-phase relative permeability curves, as the linear functions of water saturation, are provided in Figs.~\ref{fig:1}a and \ref{fig:1}b, respectively.

\begin{figure*}[ht]
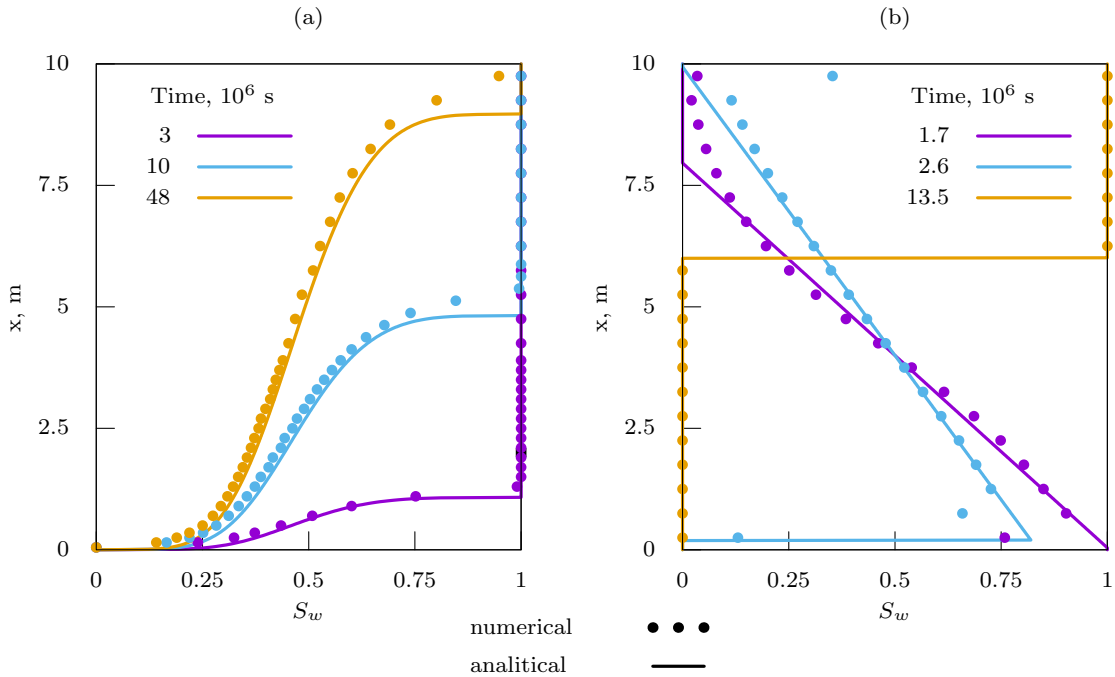
	
\centering
\input fig2.txt
\caption{Comparison between analytical and numerical solutions: (\textbf{a}) capillary imbibition; (\textbf{b}) gravity segregation.}
\label{fig:2}
\end{figure*}

The primary purpose of one-dimensional calculations is to validate and scrutinize the applied research methodology. Two-dimensional calculations are used to investigate the effectiveness and applicability of NSS-method, and analyze the convergence to numerical equilibrium of the implicit and explicit schemes. The geometrical characteristics of the one-dimensional and two-dimensional models are shown in Table~\ref{tab:1}. At the initial moment of time, the water saturation is the same for the whole thickness of all one-dimensional models and equal to $0.5$. Hence, the volume of oil and water in the reservoir is equal.

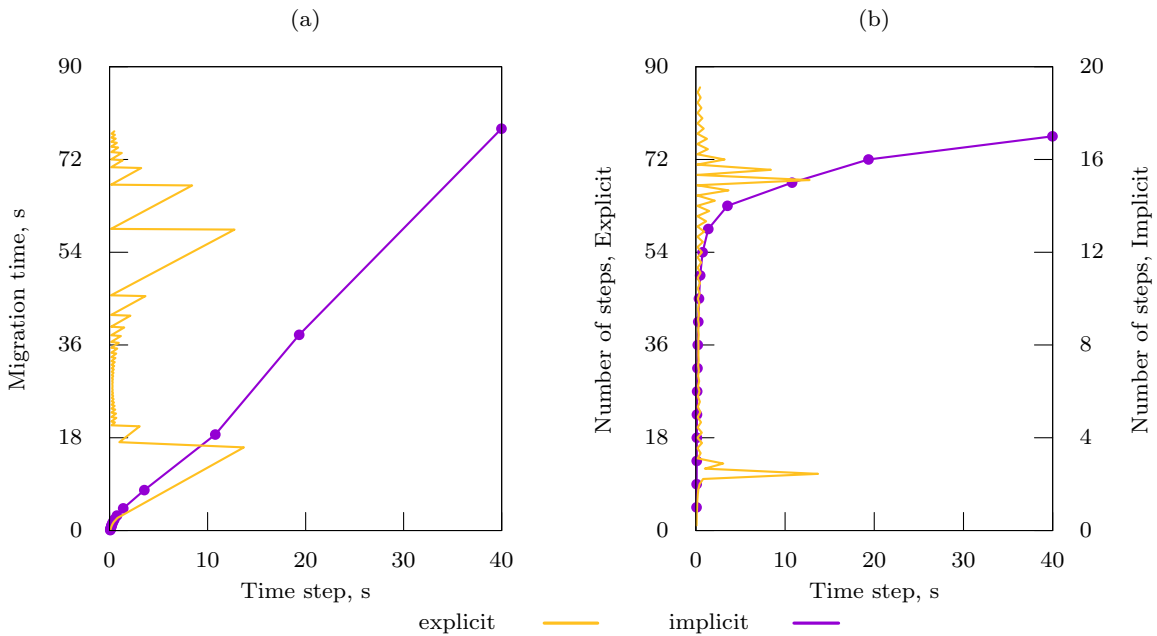
\begin{figure*}[ht]
\centering
\begin{tikzpicture}[gnuplot, scale=1.33]
\path (0.000,0.000) rectangle (11.600,5.800);
\gpcolor{color=gp lt color border}
\gpsetlinetype{gp lt border}
\gpsetlinewidth{1.00}
\draw[gp path] (1.136,0.831)--(1.316,0.831);
\node[gp node right] at (0.952,0.831) { 0};
\draw[gp path] (1.136,1.756)--(1.316,1.756);
\node[gp node right] at (0.952,1.756) { 18};
\draw[gp path] (1.136,2.681)--(1.316,2.681);
\node[gp node right] at (0.952,2.681) { 36};
\draw[gp path] (1.136,3.605)--(1.316,3.605);
\node[gp node right] at (0.952,3.605) { 54};
\draw[gp path] (1.136,4.530)--(1.316,4.530);
\node[gp node right] at (0.952,4.530) { 72};
\draw[gp path] (1.136,5.455)--(1.316,5.455);
\node[gp node right] at (0.952,5.455) { 90};
\draw[gp path] (1.136,0.831)--(1.136,1.011);
\node[gp node center] at (1.136,0.523) { 0};
\draw[gp path] (2.106,0.831)--(2.106,1.011);
\node[gp node center] at (2.106,0.523) { 10};
\draw[gp path] (3.076,0.831)--(3.076,1.011);
\node[gp node center] at (3.076,0.523) { 20};
\draw[gp path] (4.045,0.831)--(4.045,1.011);
\node[gp node center] at (4.045,0.523) { 30};
\draw[gp path] (5.015,0.831)--(5.015,1.011);
\node[gp node center] at (5.015,0.523) { 40};
\draw[gp path] (1.136,5.455)--(1.136,0.831)--(5.015,0.831)--(5.015,5.455)--cycle;
\node[gp node center,rotate=-270] at (0.246,3.143) {Migration time, s};
\node[gp node center] at (3.075,0.215) {Time step, s};
\node[gp node center] at (3.075,5.917) {(a)};
\gpcolor{rgb color={0.580,0.000,0.827}}
\gpsetlinetype{gp lt plot 0}
\gpsetlinewidth{2.00}
\draw[gp path] (1.143,0.835)--(1.144,0.839)--(1.146,0.844)--(1.147,0.850)--(1.149,0.857)%
  --(1.151,0.865)--(1.153,0.874)--(1.157,0.885)--(1.162,0.898)--(1.170,0.916)--(1.183,0.941)%
  --(1.209,0.980)--(1.271,1.052)--(1.480,1.234)--(2.182,1.788)--(3.012,2.782)--(5.013,4.837);
\gpsetpointsize{4.00}
\gppoint{gp mark 7}{(1.143,0.835)}
\gppoint{gp mark 7}{(1.144,0.839)}
\gppoint{gp mark 7}{(1.146,0.844)}
\gppoint{gp mark 7}{(1.147,0.850)}
\gppoint{gp mark 7}{(1.149,0.857)}
\gppoint{gp mark 7}{(1.151,0.865)}
\gppoint{gp mark 7}{(1.153,0.874)}
\gppoint{gp mark 7}{(1.157,0.885)}
\gppoint{gp mark 7}{(1.162,0.898)}
\gppoint{gp mark 7}{(1.170,0.916)}
\gppoint{gp mark 7}{(1.183,0.941)}
\gppoint{gp mark 7}{(1.209,0.980)}
\gppoint{gp mark 7}{(1.271,1.052)}
\gppoint{gp mark 7}{(1.480,1.234)}
\gppoint{gp mark 7}{(2.182,1.788)}
\gppoint{gp mark 7}{(3.012,2.782)}
\gppoint{gp mark 7}{(5.013,4.837)}
\gpcolor{rgb color={1.000,0.753,0.125}}
\gpsetlinetype{gp lt plot 0}
\draw[gp path] (1.143,0.835)--(1.145,0.840)--(1.146,0.845)--(1.148,0.851)--(1.150,0.859)%
  --(1.153,0.867)--(1.157,0.878)--(1.164,0.893)--(1.177,0.915)--(1.215,0.957)--(2.465,1.660)%
  --(1.233,1.712)--(1.434,1.870)--(1.156,1.880)--(1.189,1.909)--(1.154,1.918)--(1.203,1.954)%
  --(1.154,1.963)--(1.200,1.997)--(1.155,2.007)--(1.192,2.037)--(1.156,2.047)--(1.186,2.074)%
  --(1.157,2.085)--(1.180,2.108)--(1.158,2.119)--(1.176,2.141)--(1.159,2.153)--(1.172,2.172)%
  --(1.160,2.185)--(1.170,2.202)--(1.161,2.216)--(1.167,2.232)--(1.163,2.247)--(1.165,2.262)%
  --(1.165,2.277)--(1.163,2.292)--(1.167,2.308)--(1.162,2.321)--(1.169,2.339)--(1.160,2.352)%
  --(1.172,2.370)--(1.159,2.382)--(1.175,2.403)--(1.158,2.414)--(1.178,2.437)--(1.157,2.448)%
  --(1.182,2.472)--(1.156,2.483)--(1.187,2.510)--(1.155,2.520)--(1.194,2.551)--(1.154,2.560)%
  --(1.202,2.595)--(1.154,2.604)--(1.212,2.645)--(1.153,2.654)--(1.227,2.702)--(1.152,2.711)%
  --(1.248,2.770)--(1.152,2.778)--(1.281,2.855)--(1.151,2.863)--(1.342,2.973)--(1.151,2.980)%
  --(1.491,3.168)--(1.150,3.176)--(2.372,3.831)--(1.150,3.838)--(1.953,4.271)--(1.151,4.279)%
  --(1.449,4.445)--(1.152,4.453)--(1.266,4.522)--(1.152,4.531)--(1.257,4.595)--(1.153,4.604)%
  --(1.221,4.649)--(1.153,4.658)--(1.208,4.696)--(1.154,4.706)--(1.197,4.738)--(1.155,4.748)%
  --(1.189,4.776)--(1.156,4.787)--(1.183,4.811);
\gpcolor{color=gp lt color border}
\gpsetlinetype{gp lt border}
\gpsetlinewidth{1.00}
\draw[gp path] (1.136,5.455)--(1.136,0.831)--(5.015,0.831)--(5.015,5.455)--cycle;
\gpdefrectangularnode{gp plot 0}{\pgfpoint{1.136cm}{0.831cm}}{\pgfpoint{5.015cm}{5.455cm}}
\draw[gp path] (6.936,0.831)--(7.116,0.831);
\node[gp node right] at (6.752,0.831) { 0};
\draw[gp path] (6.936,1.756)--(7.116,1.756);
\node[gp node right] at (6.752,1.756) { 18};
\draw[gp path] (6.936,2.681)--(7.116,2.681);
\node[gp node right] at (6.752,2.681) { 36};
\draw[gp path] (6.936,3.605)--(7.116,3.605);
\node[gp node right] at (6.752,3.605) { 54};
\draw[gp path] (6.936,4.530)--(7.116,4.530);
\node[gp node right] at (6.752,4.530) { 72};
\draw[gp path] (6.936,5.455)--(7.116,5.455);
\node[gp node right] at (6.752,5.455) { 90};
\draw[gp path] (6.936,0.831)--(6.936,1.011);
\node[gp node center] at (6.936,0.523) { 0};
\draw[gp path] (7.819,0.831)--(7.819,1.011);
\node[gp node center] at (7.819,0.523) { 10};
\draw[gp path] (8.701,0.831)--(8.701,1.011);
\node[gp node center] at (8.701,0.523) { 20};
\draw[gp path] (9.584,0.831)--(9.584,1.011);
\node[gp node center] at (9.584,0.523) { 30};
\draw[gp path] (10.466,0.831)--(10.466,1.011);
\node[gp node center] at (10.466,0.523) { 40};
\draw[gp path] (10.466,0.831)--(10.286,0.831);
\node[gp node left] at (10.650,0.831) { 0};
\draw[gp path] (10.466,1.756)--(10.286,1.756);
\node[gp node left] at (10.650,1.756) { 4};
\draw[gp path] (10.466,2.681)--(10.286,2.681);
\node[gp node left] at (10.650,2.681) { 8};
\draw[gp path] (10.466,3.605)--(10.286,3.605);
\node[gp node left] at (10.650,3.605) { 12};
\draw[gp path] (10.466,4.530)--(10.286,4.530);
\node[gp node left] at (10.650,4.530) { 16};
\draw[gp path] (10.466,5.455)--(10.286,5.455);
\node[gp node left] at (10.650,5.455) { 20};
\draw[gp path] (6.936,5.455)--(6.936,0.831)--(10.466,0.831)--(10.466,5.455)--cycle;
\node[gp node center,rotate=-270] at (6.046,3.143) {Number of steps, Explicit};
\node[gp node center,rotate=-270] at (11.355,3.143) {Number of steps, Implicit};
\node[gp node center] at (8.701,0.215) {Time step, s};
\node[gp node center] at (8.701,5.917) {(b)};
\node[gp node left] at (4.112,-0.092) {explicit};
\node[gp node left] at (6.583,-0.092) {implicit};
\gpcolor{rgb color={1.000,0.753,0.125}}
\gpsetlinewidth{3.00}
\draw[gp path](5.436,-0.092)--(5.965,-0.092);
\gpcolor{rgb color={0.580,0.000,0.827}}
\draw[gp path](7.907,-0.092)--(8.348,-0.092);
\gpsetlinetype{gp lt plot 0}
\gpsetlinewidth{2.00}
\draw[gp path] (6.943,1.062)--(6.944,1.293)--(6.945,1.525)--(6.946,1.756)--(6.947,1.987)%
  --(6.949,2.218)--(6.952,2.449)--(6.955,2.681)--(6.960,2.912)--(6.967,3.143)--(6.979,3.374)%
  --(7.003,3.605)--(7.059,3.837)--(7.249,4.068)--(7.888,4.299)--(8.644,4.530)--(10.464,4.761);
\gppoint{gp mark 7}{(6.943,1.062)}
\gppoint{gp mark 7}{(6.944,1.293)}
\gppoint{gp mark 7}{(6.945,1.525)}
\gppoint{gp mark 7}{(6.946,1.756)}
\gppoint{gp mark 7}{(6.947,1.987)}
\gppoint{gp mark 7}{(6.949,2.218)}
\gppoint{gp mark 7}{(6.952,2.449)}
\gppoint{gp mark 7}{(6.955,2.681)}
\gppoint{gp mark 7}{(6.960,2.912)}
\gppoint{gp mark 7}{(6.967,3.143)}
\gppoint{gp mark 7}{(6.979,3.374)}
\gppoint{gp mark 7}{(7.003,3.605)}
\gppoint{gp mark 7}{(7.059,3.837)}
\gppoint{gp mark 7}{(7.249,4.068)}
\gppoint{gp mark 7}{(7.888,4.299)}
\gppoint{gp mark 7}{(8.644,4.530)}
\gppoint{gp mark 7}{(10.464,4.761)}
\gpcolor{rgb color={1.000,0.753,0.125}}
\gpsetlinetype{gp lt plot 0}
\draw[gp path] (6.943,0.882)--(6.944,0.934)--(6.945,0.985)--(6.947,1.037)--(6.949,1.088)%
  --(6.951,1.139)--(6.955,1.191)--(6.961,1.242)--(6.973,1.293)--(7.008,1.345)--(8.145,1.396)%
  --(7.025,1.448)--(7.207,1.499)--(6.954,1.550)--(6.984,1.602)--(6.952,1.653)--(6.997,1.704)%
  --(6.952,1.756)--(6.995,1.807)--(6.953,1.859)--(6.987,1.910)--(6.954,1.961)--(6.981,2.013)%
  --(6.955,2.064)--(6.976,2.115)--(6.956,2.167)--(6.972,2.218)--(6.957,2.270)--(6.969,2.321)%
  --(6.958,2.372)--(6.966,2.424)--(6.959,2.475)--(6.964,2.526)--(6.961,2.578)--(6.962,2.629)%
  --(6.962,2.681)--(6.961,2.732)--(6.964,2.783)--(6.959,2.835)--(6.966,2.886)--(6.958,2.937)%
  --(6.968,2.989)--(6.957,3.040)--(6.971,3.092)--(6.956,3.143)--(6.974,3.194)--(6.955,3.246)%
  --(6.978,3.297)--(6.954,3.349)--(6.983,3.400)--(6.953,3.451)--(6.989,3.503)--(6.953,3.554)%
  --(6.996,3.605)--(6.952,3.657)--(7.006,3.708)--(6.951,3.760)--(7.019,3.811)--(6.951,3.862)%
  --(7.038,3.914)--(6.950,3.965)--(7.068,4.016)--(6.950,4.068)--(7.124,4.119)--(6.949,4.171)%
  --(7.259,4.222)--(6.949,4.273)--(8.061,4.325)--(6.949,4.376)--(7.679,4.427)--(6.949,4.479)%
  --(7.221,4.530)--(6.950,4.582)--(7.055,4.633)--(6.950,4.684)--(7.046,4.736)--(6.951,4.787)%
  --(7.014,4.838)--(6.952,4.890)--(7.001,4.941)--(6.953,4.993)--(6.991,5.044)--(6.953,5.095)%
  --(6.984,5.147)--(6.954,5.198)--(6.978,5.249);
\gpcolor{color=gp lt color border}
\gpsetlinetype{gp lt border}
\gpsetlinewidth{1.00}
\draw[gp path] (6.936,5.455)--(6.936,0.831)--(10.466,0.831)--(10.466,5.455)--cycle;
\gpdefrectangularnode{gp plot 0}{\pgfpoint{6.936cm}{0.831cm}}{\pgfpoint{10.466cm}{5.455cm}}
\end{tikzpicture}
\caption{Time step magnitude of the explicit and implicit schemes depending on: (\textbf{a}) migration time; (\textbf{b}) number of time steps.}
\label{fig:4}       
\end{figure*}

The number of cells is reduced to one of the one-dimensional cases to analyze the stability of numerical solutions and oscillations of the explicit scheme. Consequently, the number of grid blocks is decreased ten times for a better representation of the obtained results (Fig.~\ref{fig:4}). Thus, the height of the column is modified from $10$ meters to $1$ meter.

The variable time step is implemented in both explicit and implicit schemes by specifying the time step multiplier (TSM). The time step is calculated by multiplying the maximum allowable time step by TSM, which is artificially introduced to the numerical scheme. The maximum allowable time step is obtained obeying the relation (\ref{eq:sum_sat_2ph}), and increases as the system approaches the equilibrium state and vice versa. The value of TSM is equalized for both explicit and implicit schemes for the calculations depicted in Fig.~\ref{fig:4}.

\begin{figure*}[ht]
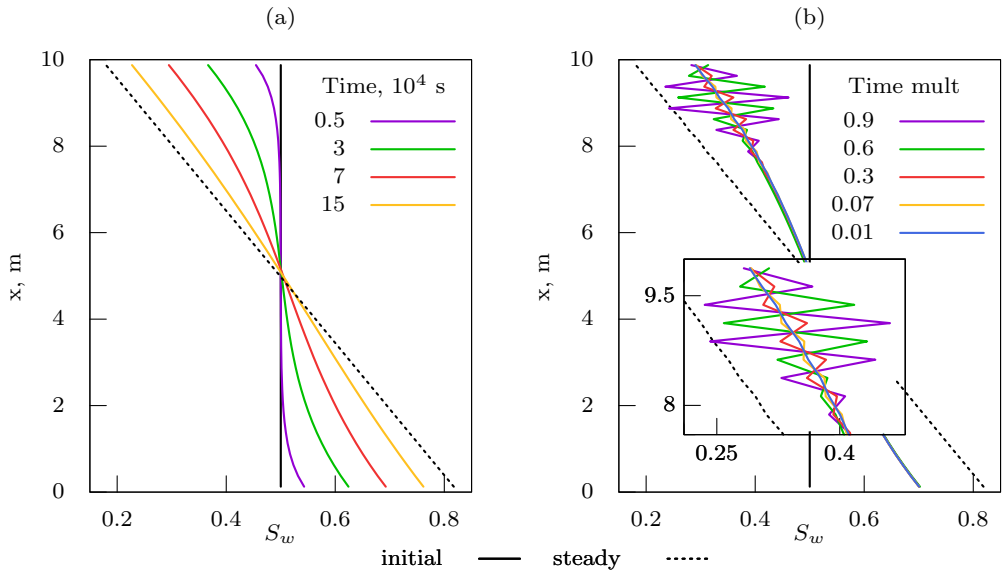

\centering
\input fig3.txt 
\caption{Capillary to gravity driven imbibition in one-dimensional case at $7 \cdot 10^{4}$ sec.: (\textbf{a}) vertical water saturation profile at different times; (\textbf{b}) oscillations of the explicit scheme depending on TSM magnitude.}
\label{fig:3}       
\end{figure*}

\begin{figure*}[ht]
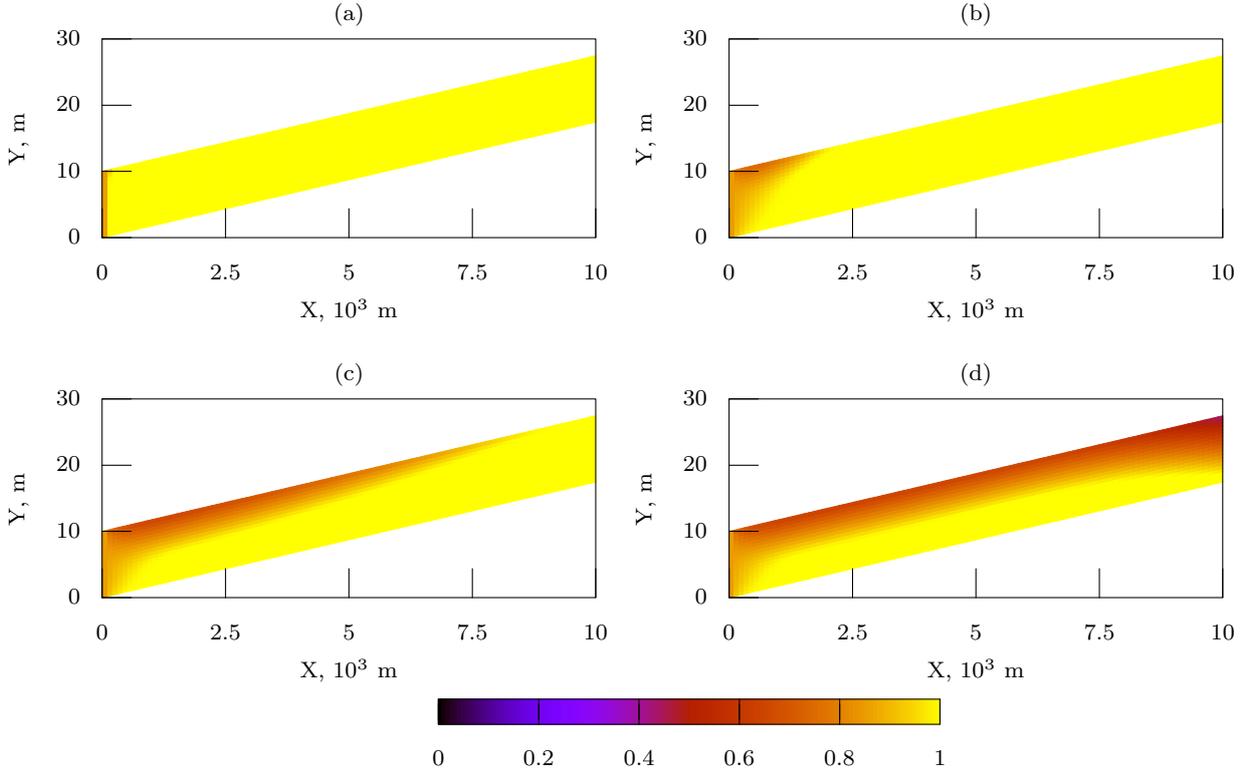

\centering
\input fig5.txt  
\caption{Oil migration (two-dimensional). Change of water saturation in the presence of an oil source: (\textbf{a})~$10^{6}$ sec.; (\textbf{b})~$10^{9}$ sec.; (\textbf{c})~$10^{10}$ sec.; (\textbf{d})~$5 \cdot 10^{10}$ sec.}
\label{fig:5}       
\end{figure*}

\begin{figure*}[ht]
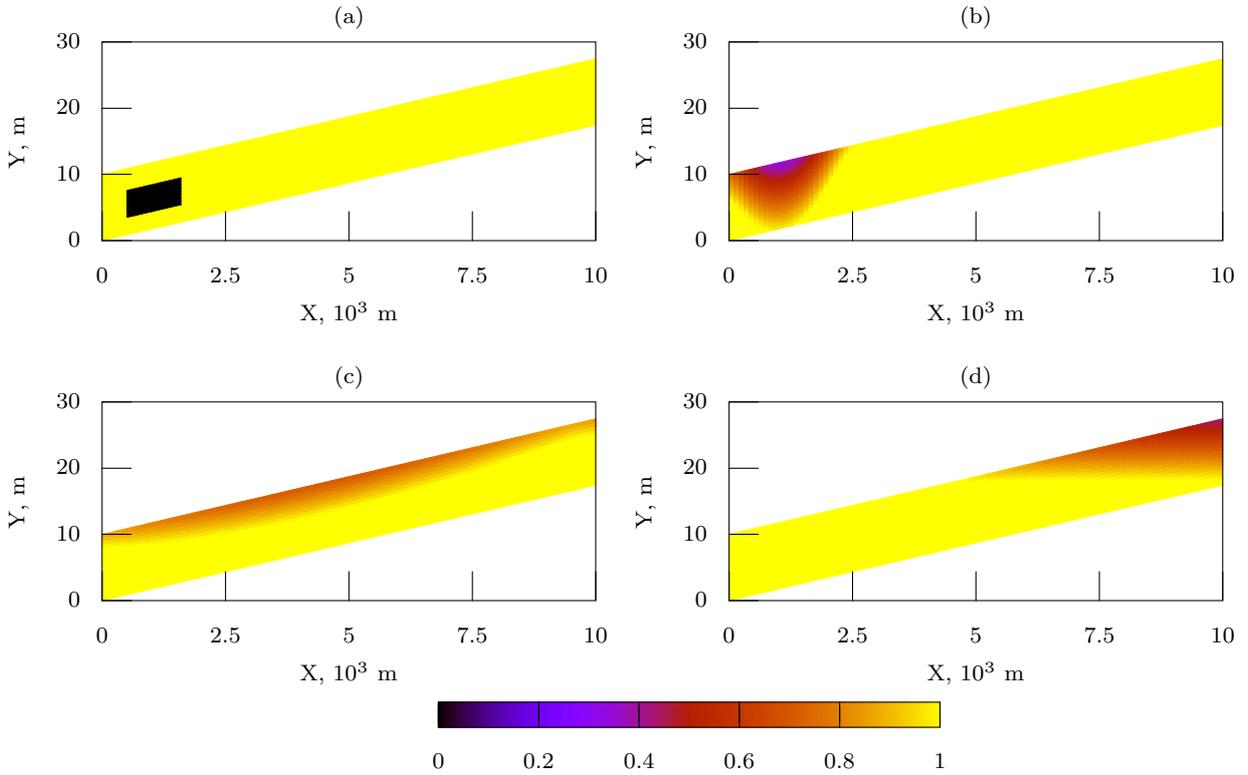

\centering
\input fig6.txt 
\caption{Oil migration (two-dimensional). Change of water saturation in the presence of uneven initial oil distribution: (\textbf{a}) 0 sec; (\textbf{b}) $10^{8}$ sec; (\textbf{c}) $10^{10}$ sec; (\textbf{d}) $5 \cdot 10^{10}$ sec.}
\label{fig:6}       
\end{figure*}

Two-dimensional calculations are illustrated in Figs.~\ref{fig:5} and \ref{fig:6}, and correspond to the cross-sectional grid system. Such calculations describe the process of oil migration in the rectangular reservoir which is inclined to the horizontal surface.

\subsection{Justification of numerical scheme splitting (NSS-method)}
\label{Dev}

Natural geological reservoirs typically possess some common geometrical characteristics. Along the sedimentary bedding, the reservoir size is several orders of magnitude greater than in the vertical direction. The influence of gravity is substantially stronger in the case of vertical flow due to the low inclination of the beds towards the horizon. Consequently, it is assumed that characteristic time required to achieve capillary-gravity equilibrium in the vertical plane is significantly lower than along the sedimentary deposition.

This approach allows solving finite-difference equations considerably reducing the computational time. Notably, the NSS-method represents a trivial modification of the conventional finite-difference method typically applied when modelling transport in porous media.  Such modification is implemented as follows: for a time step which is multiple to a certain integer number, the vertical flow is assumed to be negligible. Thus, the vertical permeability equals to zero in correspondent time step. The aforementioned integer number is further called `the degree of splitting'.

To prove the appropriate convergence of the described approach, the root-mean-square deviation (\ref{RMSdev}) of saturation is researched, showing the difference between the conventional calculation and the implemented NSS-method.

\begin{eqnarray}
\begin{gathered}
\label{RMSdev}
\sigma = \sqrt{ \frac{1}{N} \sum\limits_{cell} \left( 2 \frac{\tilde{S_{o}} - S_{o}}{\tilde{S_{o}} + S_{o}}\right)^2},
\end{gathered}
\end{eqnarray}
where $N$~--~number of cells, $\tilde{S_{o}}$~--~result of conventional simulation.

\begin{table}
\caption{Petrophysical properties} \label{tab:1}
\begin{tabular}{lll}
\hline\noalign{\smallskip}
Parameter & Value  \\
\noalign{\smallskip}\hline\noalign{\smallskip}
Porosity & $0.2$ \\
Permeability ($X$) & $10^{-12}$ $m^2$\\
Permeability ($Y$) & $10^{-13}$ $m^2$\\
Density (oil) & $800$ $kg/m^{3}$\\
Density (water) & $1000$ $kg/m^{3}$\\
Viscosity (oil) & $10^{-5}$ $Pa \cdot s$\\
Viscosity (water) & $5$ $\cdot$ $10^{-5}$ $Pa \cdot s$\\
\noalign{\smallskip}\hline
\end{tabular}
\end{table}
\hfill
\begin{table}
\caption{Geometry} \label{tab:2}
\begin{tabular}{lll}
\hline\noalign{\smallskip}
One-dimensional\\
\noalign{\smallskip}\hline\noalign{\smallskip}
Depth ($Y$) & $10$ $m$\\
Number of grid cells & $40$ \\
Block dimensions ($X, Y, Z$) & $0.5, 0.25, 1$ $m$\\
\hline\noalign{\smallskip}
Two-dimensional\\
\noalign{\smallskip}\hline\noalign{\smallskip}
Lenth ($X$) & $10^{4}$ $m$\\
Depth ($Y$) & $10$ $m$\\
Inclination & $0.1$ $deg$\\
Number of grid cells & $1881$\\
Block dimensions ($X, Y, Z$) & $100, 0.5, 1$ $m$\\
\noalign{\smallskip}\hline     
\end{tabular}
\end{table}

\subsection{Numerical schemes and NSS-method validation}
\label{Val}

\begin{figure*}[ht]
\centering
\begin{tikzpicture}[gnuplot, scale=1.5]
\path (0.000,0.000) rectangle (11.600,5.800);
\gpcolor{color=gp lt color border}
\gpsetlinetype{gp lt border}
\gpsetdashtype{gp dt solid}
\gpsetlinewidth{1.00}
\draw[gp path] (1.688,0.985)--(1.868,0.985);
\node[gp node right] at (1.504,0.985) {$0$};
\draw[gp path] (1.688,1.730)--(1.868,1.730);
\node[gp node right] at (1.504,1.730) {$0.005$};
\draw[gp path] (1.688,2.475)--(1.868,2.475);
\node[gp node right] at (1.504,2.475) {$0.01$};
\draw[gp path] (1.688,3.220)--(1.868,3.220);
\node[gp node right] at (1.504,3.220) {$0.015$};
\draw[gp path] (1.688,3.965)--(1.868,3.965);
\node[gp node right] at (1.504,3.965) {$0.02$};
\draw[gp path] (1.688,4.710)--(1.868,4.710);
\node[gp node right] at (1.504,4.710) {$0.025$};
\draw[gp path] (1.688,5.455)--(1.868,5.455);
\node[gp node right] at (1.504,5.455) {$0.03$};
\draw[gp path] (1.688,0.985)--(1.688,1.165);
\node[gp node center] at (1.688,0.677) {$2$};
\draw[gp path] (2.520,0.985)--(2.520,1.165);
\node[gp node center] at (2.520,0.677) {$7$};
\draw[gp path] (3.352,0.985)--(3.352,1.165);
\node[gp node center] at (3.352,0.677) {$12$};
\draw[gp path] (4.183,0.985)--(4.183,1.165);
\node[gp node center] at (4.183,0.677) {$17$};
\draw[gp path] (5.015,0.985)--(5.015,1.165);
\node[gp node center] at (5.015,0.677) {$22$};
\draw[gp path] (1.688,5.455)--(1.688,0.985)--(5.015,0.985)--(5.015,5.455)--cycle;
\node[gp node center,rotate=-270] at (0.246,3.220) {Deviation};
\node[gp node center] at (3.351,0.215) {Degree of splitting};
\node[gp node center] at (3.351,5.917) {(a)};
\node[gp node right] at (3.547,5.106) {$impl_{no \; source}$};
\gpcolor{rgb color={0.580,0.000,0.827}}
\gpsetlinewidth{2.00}
\draw[gp path] (3.731,5.106)--(4.647,5.106);
\draw[gp path] (1.688,1.705)--(1.854,1.571)--(2.021,1.529)--(2.187,1.506)--(2.353,1.493)%
  --(2.520,1.485)--(2.686,1.480)--(2.852,1.475)--(3.019,1.471)--(3.185,1.470)--(3.352,1.466)%
  --(3.518,1.466)--(3.684,1.463)--(3.851,1.461)--(4.017,1.463)--(4.183,1.460)--(4.350,1.461)%
  --(4.516,1.462)--(4.682,1.456)--(4.849,1.455)--(5.015,1.460);
\gpsetpointsize{4.00}
\gppoint{gp mark 7}{(1.688,1.705)}
\gppoint{gp mark 7}{(1.854,1.571)}
\gppoint{gp mark 7}{(2.021,1.529)}
\gppoint{gp mark 7}{(2.187,1.506)}
\gppoint{gp mark 7}{(2.353,1.493)}
\gppoint{gp mark 7}{(2.520,1.485)}
\gppoint{gp mark 7}{(2.686,1.480)}
\gppoint{gp mark 7}{(2.852,1.475)}
\gppoint{gp mark 7}{(3.019,1.471)}
\gppoint{gp mark 7}{(3.185,1.470)}
\gppoint{gp mark 7}{(3.352,1.466)}
\gppoint{gp mark 7}{(3.518,1.466)}
\gppoint{gp mark 7}{(3.684,1.463)}
\gppoint{gp mark 7}{(3.851,1.461)}
\gppoint{gp mark 7}{(4.017,1.463)}
\gppoint{gp mark 7}{(4.183,1.460)}
\gppoint{gp mark 7}{(4.350,1.461)}
\gppoint{gp mark 7}{(4.516,1.462)}
\gppoint{gp mark 7}{(4.682,1.456)}
\gppoint{gp mark 7}{(4.849,1.455)}
\gppoint{gp mark 7}{(5.015,1.460)}
\gppoint{gp mark 7}{(4.189,5.106)}
\gpcolor{color=gp lt color border}
\node[gp node right] at (3.547,4.769) {$impl_{with \; source}$};
\gpcolor{rgb color={0.000,0.620,0.451}}
\draw[gp path] (3.731,4.769)--(4.647,4.769);
\draw[gp path] (1.688,3.627)--(1.854,3.083)--(2.021,2.783)--(2.187,2.498)--(2.353,2.534)%
  --(2.520,2.548)--(2.686,2.557)--(2.852,2.182)--(3.019,2.255)--(3.185,2.281)--(3.352,2.290)%
  --(3.518,2.297)--(3.684,2.302)--(3.851,2.306)--(4.017,2.308)--(4.183,2.309)--(4.350,2.309)%
  --(4.516,2.308)--(4.682,2.306)--(4.849,2.304)--(5.015,2.301);
\gppoint{gp mark 7}{(1.688,3.627)}
\gppoint{gp mark 7}{(1.854,3.083)}
\gppoint{gp mark 7}{(2.021,2.783)}
\gppoint{gp mark 7}{(2.187,2.498)}
\gppoint{gp mark 7}{(2.353,2.534)}
\gppoint{gp mark 7}{(2.520,2.548)}
\gppoint{gp mark 7}{(2.686,2.557)}
\gppoint{gp mark 7}{(2.852,2.182)}
\gppoint{gp mark 7}{(3.019,2.255)}
\gppoint{gp mark 7}{(3.185,2.281)}
\gppoint{gp mark 7}{(3.352,2.290)}
\gppoint{gp mark 7}{(3.518,2.297)}
\gppoint{gp mark 7}{(3.684,2.302)}
\gppoint{gp mark 7}{(3.851,2.306)}
\gppoint{gp mark 7}{(4.017,2.308)}
\gppoint{gp mark 7}{(4.183,2.309)}
\gppoint{gp mark 7}{(4.350,2.309)}
\gppoint{gp mark 7}{(4.516,2.308)}
\gppoint{gp mark 7}{(4.682,2.306)}
\gppoint{gp mark 7}{(4.849,2.304)}
\gppoint{gp mark 7}{(5.015,2.301)}
\gppoint{gp mark 7}{(4.189,4.769)}
\gpcolor{color=gp lt color border}
\node[gp node right] at (3.547,4.432) {$expl_{no \; source}$};
\gpcolor{rgb color={0.337,0.706,0.914}}
\draw[gp path] (3.731,4.432)--(4.647,4.432);
\draw[gp path] (1.688,2.594)--(1.854,1.407)--(2.021,2.001)--(2.187,1.237)--(2.353,1.333)%
  --(2.520,1.758)--(2.686,1.220)--(2.852,1.595)--(3.019,1.203)--(3.185,1.666)--(3.352,1.207)%
  --(3.518,2.066)--(3.684,1.162)--(3.851,1.199)--(4.017,1.228)--(4.183,1.189)--(4.350,1.879)%
  --(4.516,1.890)--(4.682,1.382)--(4.849,1.223)--(5.015,1.702);
\gppoint{gp mark 5}{(1.688,2.594)}
\gppoint{gp mark 5}{(1.854,1.407)}
\gppoint{gp mark 5}{(2.021,2.001)}
\gppoint{gp mark 5}{(2.187,1.237)}
\gppoint{gp mark 5}{(2.353,1.333)}
\gppoint{gp mark 5}{(2.520,1.758)}
\gppoint{gp mark 5}{(2.686,1.220)}
\gppoint{gp mark 5}{(2.852,1.595)}
\gppoint{gp mark 5}{(3.019,1.203)}
\gppoint{gp mark 5}{(3.185,1.666)}
\gppoint{gp mark 5}{(3.352,1.207)}
\gppoint{gp mark 5}{(3.518,2.066)}
\gppoint{gp mark 5}{(3.684,1.162)}
\gppoint{gp mark 5}{(3.851,1.199)}
\gppoint{gp mark 5}{(4.017,1.228)}
\gppoint{gp mark 5}{(4.183,1.189)}
\gppoint{gp mark 5}{(4.350,1.879)}
\gppoint{gp mark 5}{(4.516,1.890)}
\gppoint{gp mark 5}{(4.682,1.382)}
\gppoint{gp mark 5}{(4.849,1.223)}
\gppoint{gp mark 5}{(5.015,1.702)}
\gppoint{gp mark 5}{(4.189,4.432)}
\gpcolor{color=gp lt color border}
\node[gp node right] at (3.547,4.095) {$expl_{with \; source}$};
\gpcolor{rgb color={0.902,0.624,0.000}}
\draw[gp path] (3.731,4.095)--(4.647,4.095);
\draw[gp path] (1.688,3.777)--(1.854,3.254)--(2.021,2.798)--(2.187,2.752)--(2.353,2.524)%
  --(2.520,2.593)--(2.686,2.341)--(2.852,2.470)--(3.019,2.319)--(3.185,2.261)--(3.352,2.198)%
  --(3.518,2.410)--(3.684,2.009)--(3.851,2.158)--(4.017,2.012)--(4.183,2.211)--(4.350,2.169)%
  --(4.516,2.014)--(4.682,1.871)--(4.849,2.046)--(5.015,1.929);
\gppoint{gp mark 5}{(1.688,3.777)}
\gppoint{gp mark 5}{(1.854,3.254)}
\gppoint{gp mark 5}{(2.021,2.798)}
\gppoint{gp mark 5}{(2.187,2.752)}
\gppoint{gp mark 5}{(2.353,2.524)}
\gppoint{gp mark 5}{(2.520,2.593)}
\gppoint{gp mark 5}{(2.686,2.341)}
\gppoint{gp mark 5}{(2.852,2.470)}
\gppoint{gp mark 5}{(3.019,2.319)}
\gppoint{gp mark 5}{(3.185,2.261)}
\gppoint{gp mark 5}{(3.352,2.198)}
\gppoint{gp mark 5}{(3.518,2.410)}
\gppoint{gp mark 5}{(3.684,2.009)}
\gppoint{gp mark 5}{(3.851,2.158)}
\gppoint{gp mark 5}{(4.017,2.012)}
\gppoint{gp mark 5}{(4.183,2.211)}
\gppoint{gp mark 5}{(4.350,2.169)}
\gppoint{gp mark 5}{(4.516,2.014)}
\gppoint{gp mark 5}{(4.682,1.871)}
\gppoint{gp mark 5}{(4.849,2.046)}
\gppoint{gp mark 5}{(5.015,1.929)}
\gppoint{gp mark 5}{(4.189,4.095)}
\gpcolor{color=gp lt color border}
\gpsetlinewidth{1.00}
\draw[gp path] (1.688,5.455)--(1.688,0.985)--(5.015,0.985)--(5.015,5.455)--cycle;
\gpdefrectangularnode{gp plot 1}{\pgfpoint{1.688cm}{0.985cm}}{\pgfpoint{5.015cm}{5.455cm}}
\draw[gp path] (7.304,0.985)--(7.484,0.985);
\node[gp node right] at (7.120,0.985) {$0$};
\draw[gp path] (7.304,2.103)--(7.484,2.103);
\node[gp node right] at (7.120,2.103) {$7.5$};
\draw[gp path] (7.304,3.220)--(7.484,3.220);
\node[gp node right] at (7.120,3.220) {$15$};
\draw[gp path] (7.304,4.338)--(7.484,4.338);
\node[gp node right] at (7.120,4.338) {$22.5$};
\draw[gp path] (7.304,5.455)--(7.484,5.455);
\node[gp node right] at (7.120,5.455) {$30$};
\draw[gp path] (7.304,0.985)--(7.304,1.165);
\node[gp node center] at (7.304,0.677) {$2$};
\draw[gp path] (8.240,0.985)--(8.240,1.165);
\node[gp node center] at (8.240,0.677) {$7$};
\draw[gp path] (9.176,0.985)--(9.176,1.165);
\node[gp node center] at (9.176,0.677) {$12$};
\draw[gp path] (10.111,0.985)--(10.111,1.165);
\node[gp node center] at (10.111,0.677) {$17$};
\draw[gp path] (11.047,0.985)--(11.047,1.165);
\node[gp node center] at (11.047,0.677) {$22$};
\draw[gp path] (7.304,5.455)--(7.304,0.985)--(11.047,0.985)--(11.047,5.455)--cycle;
\node[gp node center,rotate=-270] at (6.046,3.220) {Number of time steps, $10^{5}$};
\node[gp node center] at (9.175,0.215) {Degree of splitting};
\node[gp node center] at (9.175,5.917) {(b)};
\node[gp node center] at (8.590,5.121) {Time mult};
\node[gp node right] at (8.408,4.659) {$0.01$};
\gpcolor{rgb color={0.580,0.000,0.827}}
\gpsetlinewidth{2.00}
\draw[gp path] (8.592,4.659)--(9.508,4.659);
\draw[gp path] (7.304,1.184)--(7.491,1.340)--(7.678,1.516)--(7.865,1.706)--(8.053,1.903)%
  --(8.240,2.107)--(8.427,2.315)--(8.614,2.526)--(8.801,2.737)--(8.988,2.950)--(9.176,3.162)%
  --(9.363,3.373)--(9.550,3.583)--(9.737,3.792)--(9.924,4.000)--(10.111,4.207)--(10.298,4.412)%
  --(10.486,4.617)--(10.673,4.821)--(10.860,5.024)--(11.047,5.228);
\gpcolor{color=gp lt color border}
\node[gp node right] at (8.408,4.351) {$0.5$};
\gpcolor{rgb color={0.133,0.545,0.133}}
\draw[gp path] (8.592,4.351)--(9.508,4.351);
\draw[gp path] (7.304,0.991)--(7.491,0.995)--(7.678,1.000)--(7.865,1.006)--(8.053,1.012)%
  --(8.240,1.018)--(8.427,1.024)--(8.614,1.030)--(8.801,1.037)--(8.988,1.043)--(9.176,1.049)%
  --(9.363,1.056)--(9.550,1.062)--(9.737,1.069)--(9.924,1.075)--(10.111,1.082)--(10.298,1.088)%
  --(10.486,1.094)--(10.673,1.101)--(10.860,1.107)--(11.047,1.113);
\gpcolor{color=gp lt color border}
\gpsetlinewidth{1.00}
\draw[gp path] (7.304,5.455)--(7.304,0.985)--(11.047,0.985)--(11.047,5.455)--cycle;
\gpdefrectangularnode{gp plot 2}{\pgfpoint{7.304cm}{0.985cm}}{\pgfpoint{11.047cm}{5.455cm}}
\end{tikzpicture}
\caption{(\textbf{a}) Root-mean-square deviation of the results obtained using NSS-method from conventional ones (explicit and implicit schemes). (\textbf{b}) Time steps quantity depending on the degree of splitting. Number of time steps with no splitting: $22 \cdot 10^{5}$ with $0.5$ time step multiplier and $740 \cdot 10^{5}$ with $0.01$ time step multiplier.}
\label{fig:7}       
\end{figure*}
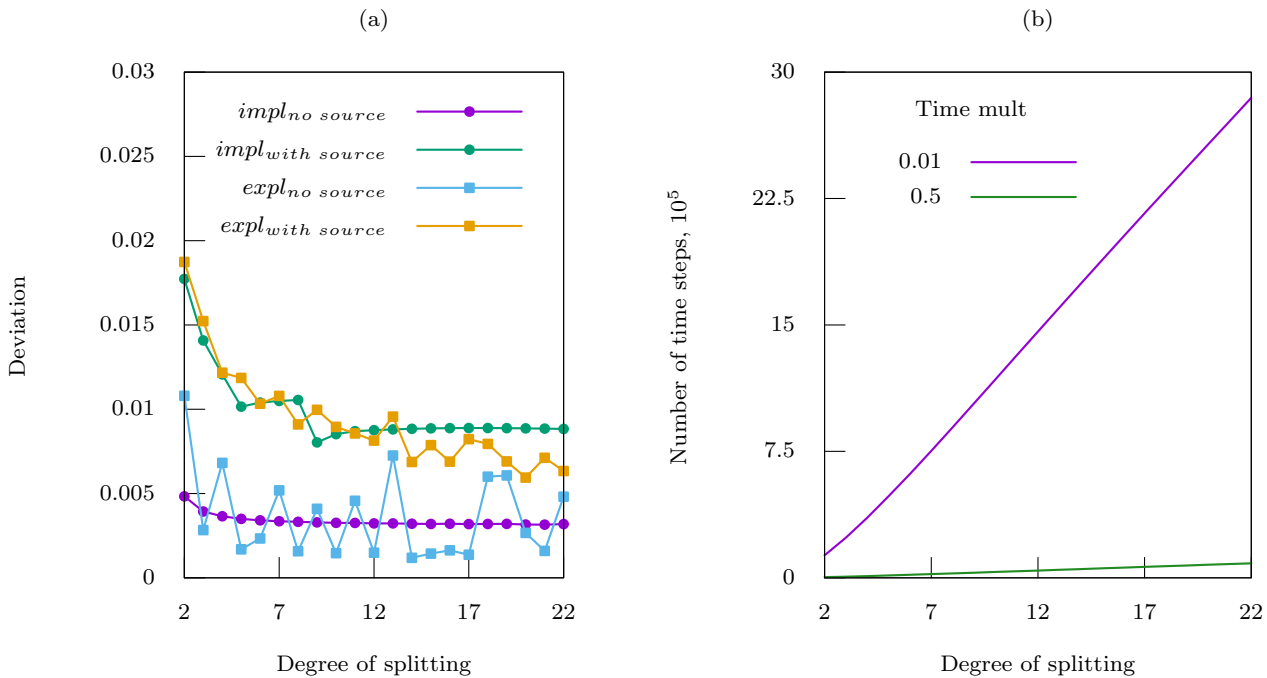

\begin{figure*}[ht]
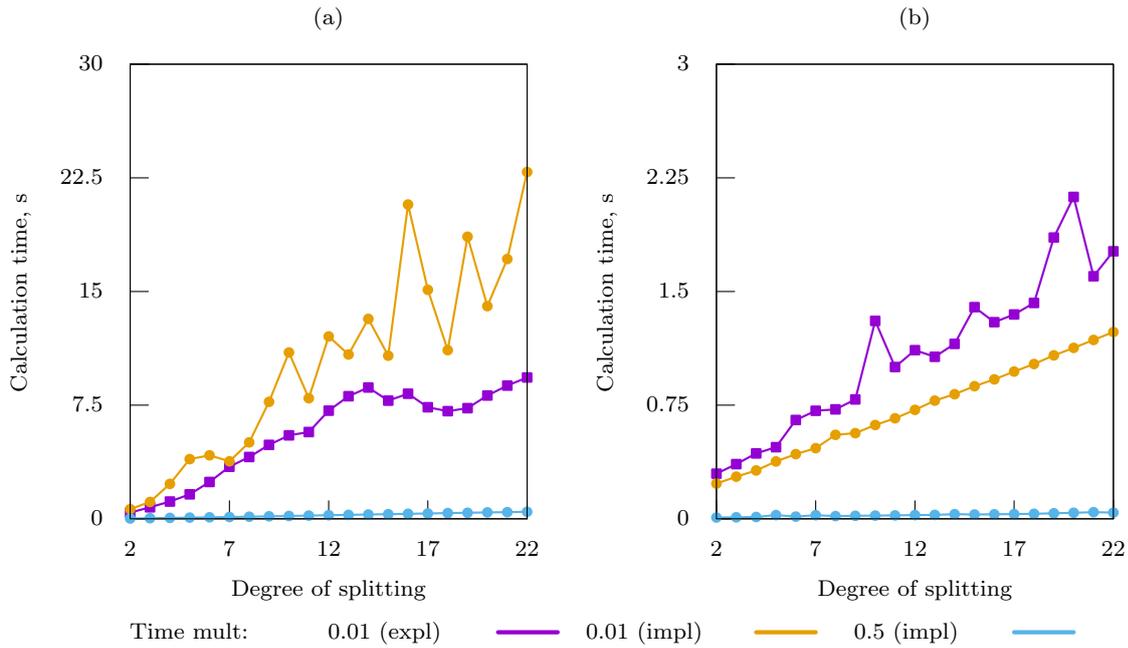

\centering
\input fig8.txt  
\caption{Computational time depending on the degree of splitting for different TSM (explicit and implicit schemes): (\textbf{a}) case with an oil source; (\textbf{b}) case with uneven initial oil distribution.}
\label{fig:8}       
\end{figure*}

The investigated NSS-method, and implemented implicit and explicit schemes are validated by comparing the results of numerical simulations with available analytical solutions. Both cases of gravity segregation and capillary imbibition have independent analytical solutions which can be found in  Mayer (2001~\cite{mayer2001gravitacionnaya}) and Kashchiev et~al~(2003~\cite{kashchiev2003analytical}), respectively (Fig.~\ref{fig:2}). Bedriko\-vetsky and Maron (1986~\cite{bedrikovetsky1986gravitacionnoie}), and Bedrikovetsky~(2013~\cite{bedrikovetsky2013mathematical}) deeply researched the analytical solutions in this branch of fluid dynamics.

Figures~\ref{fig:3} and \ref{fig:4} depict the stability of the numerical schemes. The dependence of one-dimensional solutions on the magnitude of TSM is analyzed to select the appropriate TSM value, which in this particular case equals to $0.01$ (Fig.~\ref{fig:3}). This allows achieving the appropriate calculation accuracy, despite having the numerical oscillations of the explicit scheme (Fig.~\ref{fig:4}). The effect of the proposed NSS-method on the results of two-dimensional calculations is provided in Fig.~\ref{fig:7}. The efficiency of such approach in reducing the computational time is verified in Fig.~\ref{fig:8}.

\section{Results}
\label{res}

The presented results are obtained using the numerical simulation methods described above by implementing explicit or implicit numerical schemes. Furthermore, two-dimensional calculations are conducted using NSS-method. Likewise, for the physico-mathematical model presented, there are two analytical, transient, one-dimen\-sional solutions. These two solutions are scrutinized and compared with the proposed numerical solutions (Fig.~\ref{fig:7}).

Considering the analytical solution comprising gravitational segregation, the capillary force is neglected, therefore, $P_{c} = 0$. For the sake of comparison, the relative phase permeability curves are chosen to be the same as in the implemented numerical two-dimensional solutions (Fig.~\ref{fig:1}b). Additionally, the porosity and permeability values are $0.25$ and $10 \left( md \right)$, respectively.

Regarding the analytical solution containing capillary imbibition, the capillary pressure function has a complex dependence on saturation which is described by $P_{c} = -0.4 \; ln \; S_{w} \; \left( MPa \right)$. Relative phase permeability curves can be found using the following formulas:  $k_{rw} = 0.2\; S^{4}_{w}$ and $k_{ro} = 0.25 \left(1 - S_{w}\right) ^{4}$. The porosity and permeability values are equal to $0.3$ and $20 \left( md \right)$, respectively. All the above parameters and mathematical relations are taken from  Kashchiev~et~al~(2003~~\cite{kashchiev2003analytical}).

One-dimensional calculations demonstrate the stability of the proposed numerical schemes (Fig.~\ref{fig:3}). The results of the explicit and implicit schemes coincide when a TSM is set at $0.01$ and $0.5$, respectively. Figure~\ref{fig:3}a represents the saturation profiles obtained by the implicit and explicit schemes at different moments in time. Figure~\ref{fig:3}b depicts the results obtained from the same model as in Fig.~\ref{fig:3}a, but using only an explicit scheme at time $7 \cdot 10^4$ sec., and the different values of TSM. 

The calculations presented in Fig.~\ref{fig:4} are performed with a reduced number of grid blocks in order to accelerate the  convergence to the equilibrium state. A significantly smaller number of grid blocks allows observing the numerical oscillations using an explicit scheme without averaging the results. Figures~\ref{fig:6}a and \ref{fig:4}b demonstrate the change of the time step magnitude during the calculation.

Considering two-dimensional cases, no-flow boundary conditions are used. To simulate the migration of oil, the following two methods are utilized. The first one consists of setting the oil source in blocks adjacent to the left boundary of the grid (Fig.~\ref{fig:2}). The second one defines the particular quantity of oil already existing in the reservoir water zone (Fig.~\ref{fig:3}). Such methods allow observing the imbibition process without the fluid flux through the grid boundary. The calculations for both cases are conducted using the same reservoir model.

The results of all two-dimensional calculations with the usage of NSS-method and without it are being compared. Figure~\ref{fig:7}a represents the dependence of the root-mean-square deviation on the degree of splitting. This dependence is presented for both finite-difference schemes and both cases (with oil source and without it). The relationship between the number of time steps and the degree of splitting is presented in Fig.~\ref{fig:7}b for two values of TSM $0.01$ and $0.5$, and for both numerical schemes.

Additional calculations are carried out to verify the efficiency of NSS-method. Figures~\ref{fig:8}a and \ref{fig:4} show the dependence of the calculation time on the degree of splitting. Figure~\ref{fig:8}a represents the case where the oil source is considered, while in Fig.~\ref{fig:8}b the oil source is neglected. Besides, the results with a different value of TSM are presented. The calculation times for the case accounting for the oil source and without it, with no NSS-method usage are equal to $3 \cdot 10^{2}$ $\left( sec\right)$ and $5 \cdot 10^{2}$ $\left( sec\right)$, respectively. Notably, the calculation time is comparable for the explicit and implicit schemes with TSM value $0.01$ and $0.5$, respectively. 

\section{Discussion}
\label{dis}

The obtained results, which have been listed in the previous sections, are discussed below. Since various models are compared and analyzed, the simplest fluid characteristics are chosen for convenience (Fig.~\hyperref[fig:1]{1}). The relative phase permeability functions are symmetric and have a linear dependence on saturation. However, the function $\psi$~(\ref{eq:diff_psi}) is nonlinear and non-symmetric because of the difference in water and oil viscosity (Fig.~\hyperref[fig:1]{1}b). Additionally, residual water and oil saturation are absent (Fig.~\hyperref[fig:1]{1}b), whereas the capillary pressure function has a linear form (Fig.~\ref{fig:1}a).
 
Numerical diffusion becomes vivid comparing the analytical and numerical solutions (Fig.~\ref{fig:2}). The results of both explicit and implicit solutions coincide before the curve slope changes rapidly at water saturation values around~$0.75$. The discrepancy between the results in the region of a sharp curve inflexion is a consequence of the numerical diffusion and can be minimized by reducing the size of the calculated grid blocks. Hence, these deviations between the numerical and analytical models are accepted to be appropriate.
 
Consequently, the results of one-dimensional calculations are illustrated. Figure~\ref{fig:3} shows that the implicit scheme is more stable in comparison with the explicit scheme. Notably, the explicit scheme undergoes numerical oscillations. As TSM increments, the magnitude of the oscillations increases accordingly, making the explicit scheme stable only for a small value of the time step. In our case, the numerical oscillations disappear when the value of TSM is in the range of $0.01$ or less.

The critical peculiarity distinguishing the explicit and implicit schemes for one-dimensional cases is shown in Fig.~\ref{fig:4}. The full numerical capillary gravity equilibrium in the one-dimensional reservoir can be achieved for an implicit scheme, where a time step growth is restricted only by the computational precision. Regarding the explicit scheme, the time step magnitude does not increase sufficiently while oscillating, which means that the full numerical equilibrium is not reached. One of the main aims of this research is to evaluate the possibility of the full numerical equilibrium in a two-dimensional case, utilizing the implicit scheme. The answer to this question is presented at the end of this section.

It is also essential to prove the applicability of NSS-method. Its efficiency is exemplified using two-dimensio\-nal calculations. The influence of this method on the accuracy of the calculations is shown in Fig.~\ref{fig:7}a. The dependence of the root-mean-square deviation on the degree of splitting is hyperbolic. After a certain degree of splitting, the deflection curve becomes practically parallel to the horizontal axis. The closest convergence is observed in the absence of the oil source when the explicit scheme is implemented. The value of the root-mean-square deviation is insignificant and is assumed to be acceptable for both implicit and explicit cases, either with the oil source or without it. 

Figure~\ref{fig:7}b shows how NSS-method reduces the number of time steps. If the conventional method is implemented, the number of time steps for TSM equal to $0.01$ reaches $740 \cdot 10^{5}$. As the suggested algorithm is utilized with the degree of splitting equal to $10$, the number of time steps $2 \cdot 10^{5}$ decreases approximately by $300$ times and clearly shows the method efficiency. A similar effect is observed for a TSM of $0.5$.

Figure \hyperref[fig:8]{8} shows the efficiency of NSS-method in reducing the calculation time. For instance, in the absence of the oil source and using the implicit scheme with TSM equal to $0.5$, the calculation time with no splitting is equal to $17 \cdot 10^{3}$ sec., while with NSS-method and the splitting degree of $10$, the calculation time decreases by three orders of magnitude, reaching $20$ sec.

Unlike in implicit 1D case (see Fig. \ref{fig:4}), when considering both implicit and explicit 2D cases, the time step does not increase significantly during the numerical simulations, but  rather oscillates in the close proximity to the average value. Such phenomenon occurs as the both numerical 2D schemes do not equilibrate.  

Considering the two-dimensional case (Figs. \ref{fig:5} and \ref{fig:6}), when modelling the process of migration, the full numerical equilibrium does not occur for both schemes in contrast to a one-dimensional case, where it can be reached using the implicit scheme. Thus, as it was mentioned above, the computational time remains considerably high, as the absence of numerical equilibrium does not allow increasing the time step.

It is worth noting that full numerical equilibrium can be achieved in the case when the capillary gravity segregation between the fluids inside the reservoir is completed. However, such steady condition is described by the stationary Eq.~(\ref{eq:conserv_mass_2}) (with no time-dependent first term), which has a trivial analytical solution.

\section{Conclusion}
\label{con}

The concluding section summarizes the results of the study and lists the main conclusions and findings of this paper. The main aim of this research was to ensure the applicability and operational efficiency of the developed NSS-method. Comparison of various numerical models describing migration of HCs was undertaken. Besides, the influence of the NSS-method on the numerical results was considered.

The comparison of different numerical schemes and models leads to the following conclusions:
\begin{enumerate}
\item In the case of one-dimensional calculations, when modelling a capillary gravity segregation, the implicit scheme reaches the full numerical equilibrium, while the explicit scheme does not equilibrate.
\item In the two-dimensional case, when modelling a migration process, the full numerical equilibrium is not achieved for both explicit and implicit schemes under consideration.
\item The explicit and implicit schemes can be used for migration modelling. However, application of the explicit scheme is limited due to its instability and only possible when certain criteria are picked up for particular conditions. Thus, the implicit scheme remains more preferable, since it is not sensitive to computational parameters.
\item The efficiency of NSS-method is proven for all considered two-dimensional cases in the presence or absence of oil source for both numerical schemes. The investigated method might reduce the estimated computational time up to three orders of magnitude with practically no effect on the accuracy of the final results.
\end{enumerate}

\begin{acknowledgement}
Authors are grateful to the creators of `Eigen' C++ template library for linear algebra  Guennebaud et al (2010~\cite{eigenweb}) used to implement the linear algebra algorithms. Without the support and
valuable advice of our friend Andrey Kutuzov, it would not have been possible to complete this work.
\end{acknowledgement}

\bibliographystyle{plain}
\bibliography{NSS}

\end{document}